\pdfoutput=1

\documentclass[11pt]{article}

\usepackage[]{ACL2024}

\usepackage{times}
\usepackage{latexsym}
\usepackage[T1]{fontenc}
\usepackage{natbib}

\usepackage[utf8]{inputenc}

\usepackage{multirow}
\usepackage{microtype}
\usepackage[normalem]{ulem}
\usepackage{times}
\usepackage{latexsym}
\usepackage{url}
\usepackage{balance}
\usepackage{subfigure}
\usepackage{latexsym}
\usepackage{amsmath}
\usepackage{amssymb}
\usepackage{caption}
\usepackage{graphicx}
\usepackage{url}
\usepackage{multicol}
\usepackage{multirow}
\usepackage{booktabs}
\usepackage{color}
\usepackage{subfigure}
\usepackage{array}
\usepackage{balance}
\usepackage{appendix}
\usepackage{amsmath}
\usepackage{graphicx}
\usepackage{makecell}   
\usepackage{tabularx}

\usepackage{algorithm}

\title{MARVEL: Unlocking the Multi-Modal Capability of Dense Retrieval via Visual Module Plugin}

\author{Tianshuo Zhou$^{1}$\thanks{indicates equal contribution.}, Sen Mei$^{1}$\footnotemark[1], Xinze Li$^{1}$, Zhenghao Liu$^{1}$\thanks{indicates corresponding author.}, Chenyan Xiong$^{2}$, Zhiyuan Liu$^{3}$, \\ \textbf{Yu Gu$^{1}$, Ge Yu$^{1}$}\\
$^1$Department of Computer Science and Technology, Northeastern University, China\\
$^2$Language Technologies Institute, Carnegie Mellon University, United States\\
$^3$Department of Computer Science and Technology, Institute for AI, Tsinghua University, China \\
Beijing National Research Center for Information Science and Technology, China}

\begin{document}
\maketitle

\begin{abstract}
 
This paper proposes \textbf{M}ulti-mod\textbf{A}l \textbf{R}etrieval model via \textbf{V}isual modul\textbf{E} p\textbf{L}ugin (MARVEL), which learns an embedding space for queries and multi-modal documents to conduct retrieval. MARVEL encodes queries and multi-modal documents with a unified encoder model, which helps to alleviate the modality gap between images and texts. Specifically, we enable the image understanding ability of the well-trained dense retriever, T5-ANCE, by incorporating the visual module's encoded image features as its inputs. To facilitate the multi-modal retrieval tasks, we build the ClueWeb22-MM dataset based on the ClueWeb22 dataset, which regards anchor texts as queries, and extracts the related text and image documents from anchor-linked web pages. Our experiments show that MARVEL significantly outperforms the state-of-the-art methods on the multi-modal retrieval dataset WebQA and ClueWeb22-MM. MARVEL provides an opportunity to broaden the advantages of text retrieval to the multi-modal scenario. Besides, we also illustrate that the language model has the ability to extract image semantics and partly map the image features to the input word embedding space. All codes are available at \url{https://github.com/OpenMatch/MARVEL}.

\end{abstract}
\section{Introduction}
With the growth of multimedia information on the Internet, search engines tend to return multi-modal retrieval results to better satisfy the user information need~\cite{tautkute2019deepstyle,zhu2023large}. The media information provides more vivid retrieval results, such as texts, images, videos, and more, which improves users' experiences and even changes their browsing behaviors.

\begin{figure}[t]
    \centering
    \includegraphics[width=0.9\linewidth]{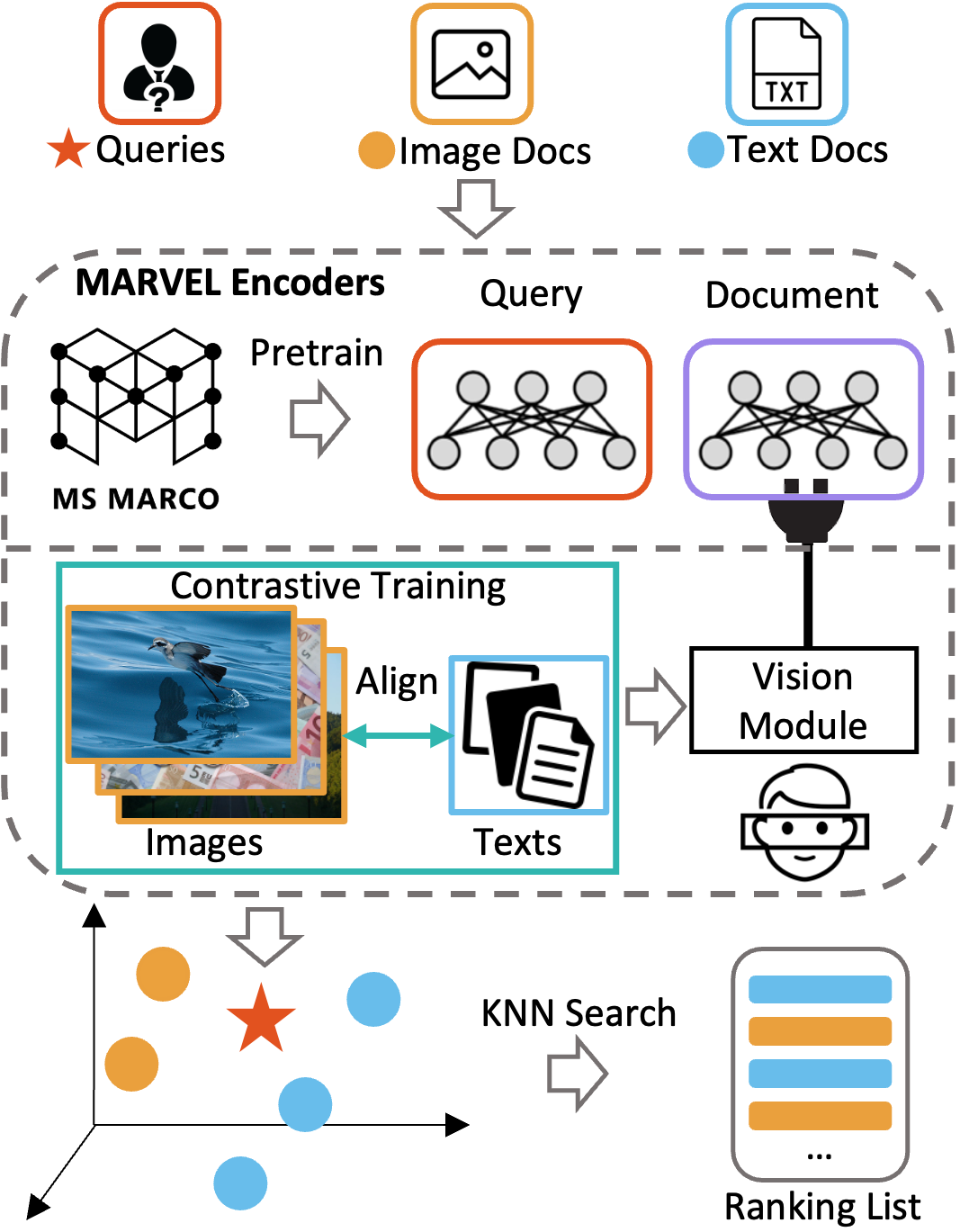}
    \caption{Retrieval Pipeline with Our MARVEL Model. MARVEL incorporates the visual module plugin, aiming to unlock the multi-modal capabilities of well trained dense retrieval model.}
    \label{fig:intro_marvel}
\end{figure}

Multi-modal retrieval~\cite{bain2021frozen,awad2021trecvid,arni2008overview,chang2022webqa} aims to return fusion results of images and texts to answer user questions. The task can be modeled using a divide-and-conquer pipeline~\cite{chang2022webqa,liu2023univldr} or universal dense retrieval~\cite{liu2023univldr}. UniVL-DR~\cite{liu2023univldr} encodes queries and multi-modal documents into a universal embedding space for multi-modal retrieval. However, this work encodes image features and texts using different encoders from CLIP~\cite{radford2021learning} and the separated text and image encoding leads to a modality gap in representing multi-modal documents. It makes UniVL-DR design an additional image verbalization method to alleviate the modality gap and also limits the text retrieval models~\cite{karpukhin2020dense,xiong2020approximate,zhan2021optimizing,li2021more,yu2021few} to excel their advantages in multi-modal scenarios.

In this paper, we propose \textbf{M}ulti-mod\textbf{A}l \textbf{R}etrieval model via \textbf{V}isual modul\textbf{E} p\textbf{L}ugin (MARVEL). As shown in Figure~\ref{fig:intro_marvel}, MARVEL is based on the text retriever T5-ANCE~\cite{yu2023openmatch}, regards the visual module as a plugin and pretrains the visual module with image-caption contrastive training for adaption. By incorporating a visual module into well-trained text retriever T5-ANCE, MARVEL seizes the opportunity to extend the benefits of unimodal learning to the multi-modal retrieval task.

To facilitate the multi-modal retrieval task, we build a large-scale benchmark, ClueWeb22-MM, based on the web page dataset, ClueWeb22~\cite{overwijk2022clueweb22}. Following previous work in text retrieval~\cite{zhang2020selective,xie2023unsupervised}, we regard the anchor text as a query and assume that its linked web page is related to the query. Subsequently, we extract image and text documents from these anchor-linked web pages. After processing, the ClueWeb22-MM encompasses over 90k queries, maintaining a scale comparable to existing benchmark WebQA~\cite{chang2022webqa}. 
Previous work~\cite{xie2023unsupervised} demonstrates that the high-quality training signals from anchor-document pairs contribute to developing a state-of-the-art dense retrieval model.

Our experiments show that MARVEL outperforms all baseline models, achieving improvements of over 2\% and 7\%, in the main metric MRR, on WebQA~\cite{chang2022webqa} and ClueWeb22-MM, respectively. The evaluation results indicate the effectiveness of MARVEL comes from the visual module plugin architecture, the visual module pretraining method, and the text matching knowledge learned by T5-ANCE. Our further analyses illustrate that the image representations encoded by the visual module can be easily captured by only finetuning the language model parameters. The training strategies guide the language model to assign more appropriate attention weights to image and text features, preventing the visual module from overfitting to the training signals. These encoded image representations not only inhabit the input embedding space for semantics alignment but also function as a kind of prompt.

\section{Related Work}
Existing dense retrieval models~\cite{karpukhin2020dense,xiong2020approximate,ren2021rocketqav2,xiong2020dense,gao2022unsupervised, luan2020sparse, khattab2020colbert} usually focus on retrieving text documents and modeling the relevance between queries and documents. They usually employ pretrained language models to encode queries and text documents into an embedding space, followed by a KNN search to retrieve candidate documents. 

Unlike the text retrieval task, the multi-modal retrieval task~\cite{chang2022webqa,hannan2020manymodalqa,singh2021mimoqa,talmor2021multimodalqa} aims to provide users with multi-modal documents that satisfy their information needs. Earlier work primarily focuses on building a divide-and-conquer pipeline for multi-modal retrieval~\cite{chang2022webqa,liu2023univldr,escalante2008late,grubinger2008overview}. In these models, retrievers individually search candidates from the document collections of different modalities and then use a reranking model to fuse the retrieval results, such as vision-language models~\cite{zhang2021vinvl}. However, this approach usually struggles to fuse the retrieval results across different modalities~\cite{liu2023univldr}. UniVL-DR~\cite{liu2023univldr} builds a universal multi-modal dense retrieval model. It encodes queries and multi-modal documents as embeddings and conducts retrieval, modality routing, and result fusion within a unified embedding space. 

Representing images is also the core of multi-modal retrieval, aiming to alleviate the modality gap between images and texts. Existing work usually focuses on representing the images using captions and image features~\cite{liu2023univldr} with different encoding methods. BERT-style pretrained visual-language models~\cite{chen2019uniter,lu2019vilbert,tan2019lxmert,su2019vl,li2019visualbert,li2021unimo,cho2021unifying,hu2020vivo,wang2022image} provide an opportunity to model the captions and image features using the same model. However, these visual-language models typically aim to align the semantics between image features and captions instead of learning representations for image documents. Thus they show less effectiveness in learning an embedding space for multi-modal retrieval~\cite{liu2023univldr}.

Another way to facilitate the image document representations is using the visual-language models that focus on representation learning, such as CLIP~\cite{radford2021learning}. It encodes image features and texts using different encoders. However, these approaches often only provide shallow interactions between texts and visual features. Thus, existing models~\cite{liu2023univldr} pay more attention to alleviating the modality gap between texts and images by the image verbalization method, aiming to bridge the modality gap between images and texts in the raw text space.

Recent advancements in multi-modal large language models~\cite{brown2020language,touvron2023llama} have introduced a novel approach to modeling multi-modality features. This approach incorporates a visual encoder module into large language models through a transformation layer~\cite{li2023blip,alayrac2022flamingo,liu2023visual}. These models extract image features using the visual encoder module of CLIP and then optimize the prompt tokens and transformation layer to map the encoded image embeddings to the raw input space of large language models~\cite{merullo2022linearly, lester2021power}. Such a visual encoder plugin method presents a unified modeling approach for handling image and text features. It not only enables the visual comprehension ability of large language models but also preserves their effectiveness by freezing their parameters.
\section{Multi-Modal Retrieval Model via Visual Module Plugin (MARVEL)}
In this section, we first describe the multi-modal retrieval (Sec.~\ref{model:mmdr}) and then introduce the model architecture of MARVEL (Sec.~\ref{model:MARVEL}). 
\begin{figure*}[t]
\centering
\subfigure[Visual Module Adaption Pretraining.] { \label{fig:model:ptetrain} 
    \includegraphics[width=0.95\linewidth]
    {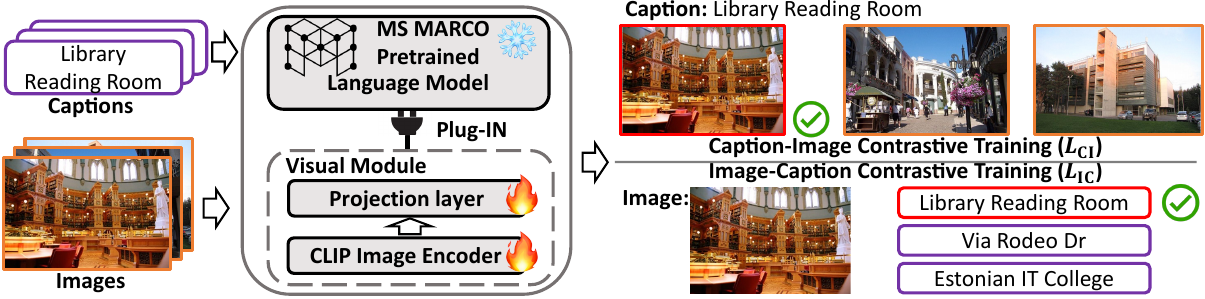}}  
\subfigure[Modality-Balanced Language Model Finetuning. We follow previous work~\cite{liu2023univldr} and sample one image document and one text document from corresponding negative document collections.] { \label{fig:model:finetune} 
    \includegraphics[width=0.95\linewidth]
    {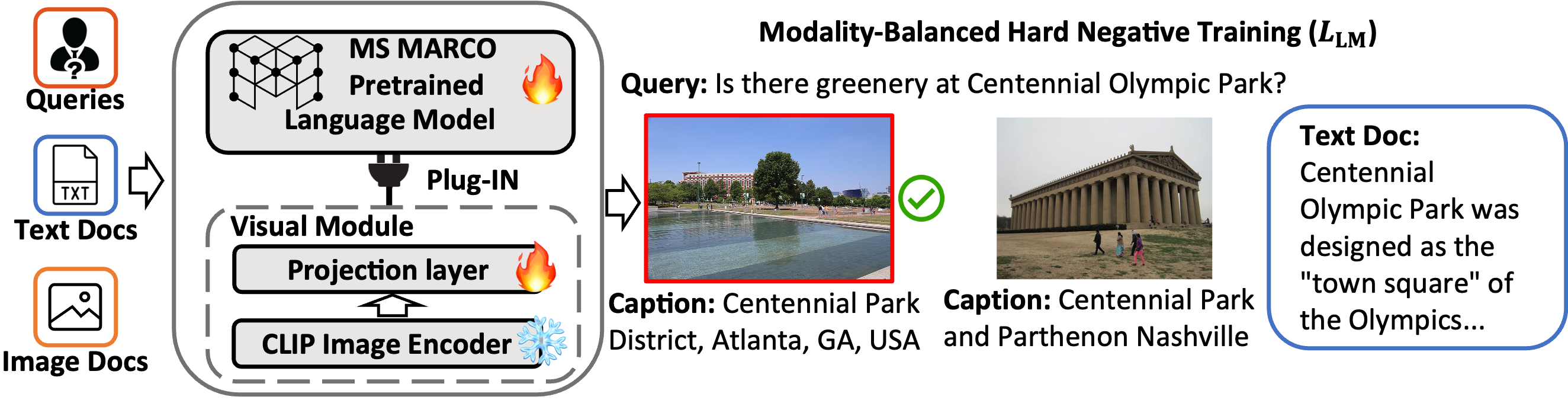}}  
\caption{The Architecture of \textbf{M}ulti-mod\textbf{A}l \textbf{R}etrieval model via \textbf{V}isual modul\textbf{E} p\textbf{L}ugin (MARVEL). We first pretrain the visual modules using the image-caption alignment task (Figure~\ref{fig:model:ptetrain}) and then finetune the language model to conduct multi-modal retrieval (Figure~\ref{fig:model:finetune}).}
\label{fig:model}
\end{figure*}

\subsection{Preliminary of Multi-Modal Retrieval}~\label{model:mmdr}
Given a query $q$, the retrieval task requires the dense retrieval models to search relevant documents from the document collection $\mathcal{D}$ to meet the information needs of users.

Previous dense retrieval models~\cite{karpukhin2020dense,xiong2020approximate,gao2021condenser,yu2023openmatch} usually focus on the text retrieval task, which aims to model the relevance between user query $q$ and text documents 
$\mathcal{D} = \{d_\text{Text}^1, ..., d_\text{Text}^m\}$. They encode both query and the $i$-th document $d_\text{Text}^i$ using language models, such as BERT~\cite{devlin2018bert}, RoBERTa~\cite{liu2019roberta} and T5~\cite{raffel2020exploring}:
\begin{equation}\label{eq:textencode}
\small
\vec{q} = \text{TextEncoder}(q); \vec{d}_\text{Text}^i = \text{TextEncoder}(d_\text{Text}^i).
\end{equation}
Different from text retrieval~\cite{nguyen2016ms,thakur2021beir}, the multi-modal retrieval task~\cite{chang2022webqa} aims to return a fusion result of documents from the collection $\mathcal{D}$, which are from different modalities. The document collection $\mathcal{D}$ not only contains texts $\mathcal{T} = \{d_\text{Text}^1, ..., d_\text{Text}^m\}$, but also includes images $\mathcal{I} = \{d_\text{Image}^1, ..., d_\text{Image}^n\}$. 

The multi-modal retrieval task requires retrievers to conduct relevance modeling, cross-modal matching, and modality fusion~\cite{liu2023univldr}. Previous work~\cite{liu2023univldr} maps text and image documents in an embedding space for retrieval, encodes texts and images using different encoders, and tries to bridge the modality gap using image verbalization methods. However, this limits the capability of dense retrieval models, hindering the expansion of text matching knowledge for learning representations for multi-modal documents.

\subsection{Universal Multi-Modal Encoding}~\label{model:MARVEL}
We show the model architecture in Figure~\ref{fig:model}. Different from previous work~\cite{liu2023univldr}, we can universally encode query $q$ and multi-modal documents $\mathcal{D} = \{d_\text{Text}^1, ..., d_\text{Text}^m, d_\text{Image}^1, ..., d_\text{Image}^n\}$ using one encoder, T5-ANCE-CLIP:
\begin{equation}
\small
  \begin{split}
    \vec{q} &= \text{T5-ANCE-CLIP}(q); \\
\vec{d}_\text{Text}^i &= \text{T5-ANCE-CLIP}(d_\text{Text}^i); \\
\vec{d}_\text{Image}^i &= \text{T5-ANCE-CLIP}(d_\text{Image}^i(I), d_\text{Image}^i(C)),
\end{split}
\end{equation}
where $d_\text{Image}^i(I)$ and $d_\text{Image}^i(C)$ are the image feature and caption of the $i$-th image document $d_\text{Image}^i$.

Then we calculate the relevance score $f(q, d^i)$ between query $q$ and the $i$-th document $d^i$ using cosine similarity:
\begin{equation}
\small
    f(q, d^i) = \cos (\vec{q}, \vec{d}^i).
\end{equation}
Following this, we conduct KNN search~\cite{johnson2019billion} to retrieve multi-modal document candidates for the given query $q$.

Subsequently, we first introduce the visual module plugin architecture of our MARVEL model (Sec.~\ref{sec:plugin}). Then we adapt the visual module to T5-ANCE by pretraining the visual understanding module (Sec.~\ref{sec:pretrain}). Finally, we finetune the parameters of T5-ANCE to learn an embedding space for multi-modal retrieval (Sec.~\ref{sec:finetune}).

\subsubsection{Dense Retrieval with Visual Plugin}\label{sec:plugin}
MARVEL starts from the T5-ANCE model~\cite{yu2023openmatch}, which is a dense retrieval model that is well-trained using MS MARCO dataset~\cite{nguyen2016ms}. Then we enable T5-ANCE by incorporating the visual module from the vision-language model, CLIP~\cite{radford2021learning}, and conduct the T5-ANCE-CLIP model. We can use a universal encoder, T5-ANCE-CLIP, to encode texts, image features, and image documents. 

Specifically, we encode the image feature $I$ using the visual encoder of CLIP~\cite{radford2021learning} and get its encoded visual representation $\vec{h}^I$:
\begin{equation}\label{eq:clip}
\small
\vec{h}^I = \text{CLIP}(I),
\end{equation}
This representation is obtained from the grid features of the last layer of the visual encoder of CLIP, and $\vec{h}^I=\{\vec{h}^I_1, ..., \vec{h}^I_{49}\}$. 
Here 49 is the number of patches. Then we follow the previous visual-language model~\cite{merullo2022linearly} and use a linear transformation layer to adapt the visual features $\vec{h}^I_{i}$ into the embedding space of the inputs of dense retrieval model:
\begin{equation}\label{eq:linear}
\small
\vec{I}_i = \text{Linear}(\vec{h}^I_{i}).
\end{equation}
Finally, we can feed these encoded image features $\vec{I} = \{\vec{I}_1, ..., \vec{I}_{49}\}$ as the ahead input embeddings $\vec{X}$ for T5-ANCE:
\begin{equation}\label{eq:imagemap}
\small
\vec{X} = \vec{e}(\text{<start>}); \vec{I}_1; ... ;\vec{I}_{49}; \vec{e}(\text{<end>}); \vec{e}_1;...; \vec{e}_k,
\end{equation}
where $;$ is the concatenation operation. $\vec{e}(\text{<start>})$ and $\vec{e}(\text{<end>})$ are the embeddings of prompt tokens to denote the start and end of encoded image feature representations. $\{\vec{e}_1...; \vec{e}_k\}$ are the word embeddings of the text input sequence $T=\{T_1,...,T_k\}$.

Different from these visual-language models~\cite{alayrac2022flamingo,li2023blip,liu2023visual,tsimpoukelli2021multimodal}, our MARVEL model aims to bring the advance of text retrieval-based pretraining to multi-modal retrieval tasks by using the visual model plugin to bridge the modality gap between images and texts.

\subsubsection{Visual Module Adaption Pretraining}\label{sec:pretrain}
In MARVEL, we adapt the visual understanding module to T5-ANCE by only pretraining the parameters of the visual module (Eq.~\ref{eq:clip}) and the projection layer (Eq.~\ref{eq:linear}). We follow~\citet{radford2021learning} and leverage the image-caption contrastive training loss $L_{\text{VM}}$ to pretrain the visual understanding module. The training loss utilizes the alignment between image features $I$ and captions $C$:
\begin{equation}
\small
L_{\text{VM}}= L_{\text{IC}} + L_{\text{CI}},
\end{equation}
where $L_{\text{IC}}$ and $L_{\text{CI}}$ are the dual direction training losses to regard image and caption as queries and then map them with corresponding caption and image, respectively:
\begin{equation}
 \small
  L_{\text{IC}} = -\log\frac{e^{f(I,C^+)/\tau}}
    {e^{f(I,C^+)/\tau} + \sum_{C^-\in \mathcal{D}_C^-}{e^{f(I,C^-)/\tau}}},
\end{equation}
\begin{equation}
 \small
  L_{\text{CI}} = -\log\frac{e^{f(C,I^+)/\tau}}
    {e^{f(C,I^+)/\tau} + \sum_{I^-\in \mathcal{D}_I^-}{e^{f(C,I^-)/\tau}}},
\end{equation}
where $\tau$ is the temperature used to scale the similarity score. $\mathcal{D}_C^-$ and $\mathcal{D}_I^-$ contain negative captions and negative images respectively, which are sampled from in-batch negatives.

\subsubsection{Modality-Balanced Language Model Finetuning}\label{sec:finetune}
During finetuning, we can freeze the parameters of the visual module (Eq.~\ref{eq:clip}) and optimize other parameters of MARVEL. To enable the MARVEL model to learn a universal embedding space for both queries and multi-modal documents, we follow previous work~\cite{liu2023univldr} and employ modality-balanced hard negative training to alleviate the modality discrimination of retrieval models:
\begin{equation}
\small
  \begin{split}
\label{eq:ce_train}
\hspace{-2mm}
 & L_{\text{LM}} = -\log\frac{e^{f(q,d^+)/\tau}}
    {e^{f(q,d^+)/\tau} + \sum_{d^-\in \mathcal{D}^-}{e^{f(q,d^-)/\tau}}} \\
    &\propto - \underbrace{f(q,d^+)/\tau}_{L_\text{Align}} + \log (\sum_{d^-\in \mathcal{D}^-}(\underbrace{e^{f(q,d_{\,\text{Image}}^-)/\tau}}_{L_\text{Image}} + \underbrace {e^{f(q,d_{\,\text{Text}}^-)/\tau}}_{L_\text{Text}})),
    \end{split}
\end{equation}
where $\mathcal{D}^-$ contains the same number of negative documents of image and text. $L_\text{Align}$ teaches models to align the query with related documents. $L_\text{Text}$ and $L_\text{Image}$ guide retrievers to choose the modality and make the embedding space uniform~\cite{liu2023univldr,wang2020understanding,chen2020simple}.
\section{Experimental Methodology}\label{sec:exp}
This section describes datasets, evaluation metrics, baselines and implementation details.

\begin{table}
\centering

\resizebox{\linewidth}{!}{%
\begin{tabular}{l|l  |  r| r r r }
\hline  
\multirow{2}{*}{\textbf{Dataset}} & \multirow{2}{*}{\textbf{Modality}} & \multirow{2}{*}{\textbf{\#Doc}} & \multicolumn{3}{c}{\textbf{\#Query}} \\
& & &{Train} & {Dev} & {Test}\\ \hline
\multirow{3}{*}{\textbf{WebQA}} & Image & 389,750 & 16,400 & 2,554 & 2,511  \\
&Text & 787,697 & 15,366 & 2,446 & 2,455   \\
&Multi-Modal & 1,177,447 & 31,766 & 5,000 & 4,966   \\
\hline

\multirow{3}{*}{\textbf{ClueWeb22-MM}} & Image & 368,710 & 35,873 & 5,041 & 5,030  \\
&Text & 363,508 & 36,155 & 4,959 & 4,970   \\
&Multi-Modal & 732,218 & 72,028 &  10,000 &  10,000   \\\hline
\end{tabular}}
\caption{\label{tab:dataset}Data Statistics.}
\end{table}
\textbf{Dataset.}
During pretraining, we collect the image-caption pairs from ClueWeb22~\cite{overwijk2022clueweb22} to train the visual understanding module. More details of pretraining data are shown in Appendix~\ref{app:pretrain}. Then two multi-modal retrieval datasets, WebQA and ClueWeb22-MM, are used for finetuning and evaluation. The data statistics are shown in Table~\ref{tab:dataset}.

WebQA is a multi-hop, multi-modal, open-domain question answering benchmark~\citep{chang2022webqa}. The dataset contains images and passage snippets that are crawled from the general Web and Wikipedia. We follow previous work~\cite{liu2023univldr} to keep the same experimental settings to preprocess the dataset. Besides, we build a new multi-modal retrieval dataset, ClueWeb22-MM, based on ClueWeb22~\cite{overwijk2022clueweb22}, which provides 10 billion web pages with rich information. We only retain web pages in English and build the ClueWeb22-MM dataset. We establish query-document relations by pairing anchors with their corresponding document~\cite{xie2023unsupervised,zhang2020selective}. 
And then we regard the anchor texts as queries and extract image documents and text documents from the linked documents. More details of building the ClueWeb22-MM dataset are shown in Appendix~\ref{app:clueweb}.

\textbf{Evaluation Metrics.} We use NDCG@10, MRR@10 and Recall@100 as evaluation metrics. Following previous work ~\cite{liu2023univldr,nguyen2016ms}, we regard MRR@10 as our main evaluation. MRR and NDCG are computed using the official scripts\footnote{\url{https://github.com/microsoft/MSMARCO-Passage-Ranking/blob/master/ms_marco_eval.py}}. Statistic significances are tested by permutation test with P$<0.05$.

\textbf{Baselines.} 
In our experiments, we follow previous work~\cite{liu2023univldr} to conduct baseline models and divide them into three groups: single modality retrieval, divide-and-conquer, and universal dense retrieval models.

\textit{Single Modality Retrieval.} In our experiments, we represent image documents using captions and use several text retrieval models as baselines. BM25~\cite{robertson2009probabilistic} is widely used in text retrieval work, which conducts exact matches between queries and documents. DPR~\cite{karpukhin2020dense} is trained using NQ dataset~\cite{kwiatkowski2019natural} and uses a dual-encoder to encode queries and documents as dense vectors for retrieval. We start from vanilla BERT~\cite{devlin2018bert} and DPR~\cite{karpukhin2020dense} checkpoints and train the encoders using in-batch negatives to conduct BERT-DPR and NQ-DPR models. NQ-ANCE is also compared, which continuously trains NQ-DPR using hard negatives~\cite{xiong2020approximate}. Besides, T5-ANCE~\cite{yu2023openmatch} and Anchor-DR~\cite{xie2023unsupervised} are dense retrieval models that are trained on MS MARCO and ClueWeb22, respectively.

\textit{Divide-and-Conquer.} The divide-and-conquer models retrieve image documents and text documents individually and then fuse the retrieval results. Following previous work~\cite{liu2023univldr}, we use single modality retrievers, VinVL-DPR, CLIP-DPR and BM25, and fuse the retrieval results according to their unimodal rank reciprocals.

\begin{table*}[ht]
  \centering
\small
  \resizebox{\linewidth}{!}{
    \begin{tabular}{l|l|ccc|ccc}
    \hline
    \multirow{2}{*}{\textbf{Setting}} & \multirow{2}{*}{\textbf{Model}} & \multicolumn{3}{c|}{\textbf{WebQA}} & \multicolumn{3}{c}{\textbf{ClueWeb22-MM}}\\\cline{3-8}
    & &  \textbf{MRR@10} &\textbf{NDCG@10}  &  \textbf{Rec@100}  & \textbf{MRR@10}  &  \textbf{NDCG@10} & \textbf{Rec@100}\\ 
    \hline
     \multirow{8}{*}{\shortstack{Single Modality\\(Text Only)}}& BM25  & 53.75 & 49.60 & 80.69 & 40.81 & 46.08 & 78.22\\
     & DPR (Zero-Shot) & 22.72 & 20.06 & 45.43 & 20.59 & 23.24 & 44.93 \\
     & CLIP-Text (Zero-Shot)& 18.16 & 16.76  & 39.83 & 30.13 & 33.91 & 59.53\\
     & Anchor-DR (Zero-Shot) & 39.96 & 37.09 & 71.32 & 42.92 & 48.50& 76.52 \\
     & T5-ANCE (Zero-Shot) & 41.57 & 37.92 & 69.33 & 45.65 & 51.71 & 83.23 \\
    & BERT-DPR & 42.16 & 39.57 & 77.10 & 38.56 & 44.41 & 80.38\\
    & NQ-DPR &  41.88 & 39.65 & 78.57 & 39.86 & 46.15 & 83.50\\ 
    & NQ-ANCE &  45.54 & 42.05 & 69.31 & 45.89 & 51.83 & 81.21\\ \hline
    \multirow{3}{*}{Divide-Conquer} & VinVL-DPR & 22.11 & 22.92 & 62.82 & 29.97 & 36.13 & 74.56 \\
    &CLIP-DPR & 37.35 & 37.56 & 85.53 & 39.54 & 47.16 & 87.25 \\
    & BM25 \& CLIP-DPR & 42.27 & 41.58 & 87.50 & 41.58 & 48.67 & 83.50 \\\hline
    \multirow{6}{*}{UnivSearch} & CLIP (Zero-Shot) & 10.59 & 8.69 & 20.21 & 16.28 & 18.52 & 40.36 \\ 
    & VinVL-DPR & 38.14 & 35.43 & 69.42 & 35.09 & 40.36 & 75.06 \\
    & CLIP-DPR & 48.83$\text{}^{}$
 & 46.32$\text{}^{}$ & 86.43$\text{}^{}$ & 42.59$\text{}^{}$
 & 49.24$\text{}^{}$
 & 87.07$\text{}^{}$
 \\
    & UniVL-DR & \uline{62.40}\rlap{$\text{}^{\dagger \mathsection}$}
 & \uline{59.32}\rlap{$\text{}^{\dagger \mathsection}$} & \uline{89.42}\rlap{$\text{}^{\dagger \mathsection}$} & 
 \uline{47.99}\rlap{$\text{}^{\dagger \mathsection}$}
 & \uline{55.41}\rlap{$\text{}^{\dagger \mathsection}$}
 & \uline{90.46}\rlap{$\text{}^{\dagger \mathsection}$}
 \\
    & MARVEL-DPR & 55.71\rlap{$\text{}^{\dagger}$}
 & 52.94\rlap{$\text{}^{\dagger}$} & 88.23\rlap{$\text{}^{\dagger}$} & 46.93\rlap{$\text{}^{\dagger}$}
 & 53.76\rlap{$\text{}^{\dagger}$}
 & 88.74\rlap{$\text{}^{\dagger}$}
 \\
    & MARVEL-ANCE & \textbf{65.15}\rlap{$\text{}^{\dagger \ddagger \mathsection}$} & \textbf{62.95}\rlap{$\text{}^{\dagger \ddagger \mathsection}$} & \textbf{92.40}\rlap{$\text{}^{\dagger \ddagger \mathsection}$} & \textbf{55.19}\rlap{$\text{}^{\dagger \ddagger \mathsection}$} & \textbf{62.83}\rlap{$\text{}^{\dagger \ddagger \mathsection}$} & \textbf{93.16}\rlap{$\text{}^{\dagger \ddagger \mathsection}$} \\\hline
    
    \end{tabular}}
    \caption{\label{tab:overall}Overall Performance. We keep the same experimental settings with previous work~\cite{liu2023univldr}. ${\dagger}$, ${\ddagger}$ and ${\mathsection}$ indicate statistically significant improvements over $\text{CLIP-DPR}^{\dagger}$, $\text{UniVL-DR}^{\ddagger}$ and $\text{MARVEL-DPR}^{\mathsection}$.}
\end{table*}
\textit{Universal Dense Retrieval.} CLIP-DPR and VinVL-DPR employ the visual language models CLIP~\cite{radford2021learning} and VinVL~\cite{zhang2021vinvl} as image and text encoders and then are trained with in-batch negatives. UniVL-DR~\cite{liu2023univldr} is our main baseline model, which further uses modality-balanced hard negative to train text and image encoders and also utilizes the image verbalization method to bridge the modality gap between images and texts.

\textbf{Implementation Details.} In our experiments, we use T5-ANCE~\cite{yu2023openmatch} as our backbone language model, which is well-trained on the MS MARCO dataset~\cite{nguyen2016ms}. Then we implement our MARVEL model by utilizing CLIP as the visual understanding module to empower the image understanding capability of T5-ANCE. The visual encoder is initialized with the clip-vit-base-patch32 checkpoint from OpenAI\footnote{\url{https://github.com/openai/CLIP}}. For MARVEL, we truncate queries, text documents and image captions to 128 tokens and set the max number of visual tokens to 49. 

During training, we use AdamW~\cite{loshchilov2018decoupled} optimizer and set maximum training epoch=20, batch size=64, learning rate=$5e-6$, and the temperature hyperparameter $\tau=0.01$. We follow UniVL-DR~\cite{liu2023univldr} and conduct MARVEL-ANCE by starting from in-batch negative finetuned MARVEL-DPR, and continuously training MARVEL-DPR with modality-balanced hard negatives. These hard negatives are randomly sampled from the top 100 retrieved negatives using MARVEL-DPR. All models are evaluated per 500 steps and the early stop step is set to 5.

\begin{table*}[ht]
  \centering
  \resizebox{\linewidth}{!}{
    \begin{tabular}{l|l|ccc|ccc}
    \hline
    \multirow{2}{*}{\textbf{Model}} & \multirow{2}{*}{\textbf{Modality}} & \multicolumn{3}{c|}{\textbf{WebQA}} & \multicolumn{3}{c}{\textbf{ClueWeb22-MM}}\\\cline{3-8}
    & &  \textbf{MRR@10} &\textbf{NDCG@10}  &  \textbf{Rec@100}  & \textbf{MRR@10}  &  \textbf{NDCG@10} & \textbf{Rec@100}\\ 
    \hline
    \multirow{3}{*}{\shortstack{MARVEL-ANCE}}& Text  & 64.72\rlap{$\text{}^{\ddagger}$} & 58.88\rlap{$\text{}^{\ddagger \mathsection}$} & 90.26\rlap{$\text{}^{\ddagger \mathsection}$} & 71.73\rlap{$\text{}^{\dagger \ddagger \mathsection}$} & 75.40\rlap{$\text{}^{\dagger \ddagger \mathsection}$} & 92.29\rlap{$\text{}^{\ddagger \mathsection}$}\\
     & Image & 66.12\rlap{$\text{}^{\dagger}$} & 67.49\rlap{$\text{}^{\dagger \ddagger}$} & 95.12\rlap{$\text{}^{\dagger \ddagger \mathsection}$} & 77.57\rlap{$\text{}^{\dagger \ddagger \mathsection}$} & 81.34\rlap{$\text{}^{\dagger \ddagger \mathsection}$} & 96.50\rlap{$\text{}^{\dagger \ddagger}$} \\
     & Multi& 65.15\rlap{$\text{}^{\ddagger}$} & 62.95\rlap{$\text{}^{\dagger \ddagger}$}  & 92.40\rlap{$\text{}^{\ddagger \mathsection}$} & 55.19\rlap{$\text{}^{\ddagger \mathsection}$} & 62.83\rlap{$\text{}^{\ddagger \mathsection}$} & 93.16\rlap{$\text{}^{\ddagger \mathsection}$}\\\hline
     \multirow{3}{*}{\shortstack{w/o CLIP Pretraining}}& Text  & 64.63\rlap{$\text{}^{\ddagger}$}  & 58.79\rlap{$\text{}^{\ddagger}$}  & 90.21\rlap{$\text{}^{\ddagger \mathsection}$}  & 70.92\rlap{$\text{}^{\mathsection}$} & 74.67\rlap{$\text{}^{\mathsection}$} & 92.13\rlap{$\text{}^{\ddagger \mathsection}$} \\
     & Image & 65.17 & 66.69 & 94.64 & 76.99\rlap{$\text{}^{\ddagger \mathsection}$} & 80.83\rlap{$\text{}^{\ddagger \mathsection}$} & 96.22 \\
     & Multi& 64.66 & 62.50\rlap{$\text{}^{\ddagger}$}  & 92.24\rlap{ $\text{}^{\ddagger \mathsection}$} & 55.18\rlap{$\text{}^{\ddagger \mathsection}$} & 62.81\rlap{$\text{}^{\ddagger \mathsection}$} & 93.07\rlap{$\text{}^{\ddagger}$}\\\hline
      \multirow{3}{*}{\shortstack{w/o MS MARCO Pretraining}}& Text  & 63.37 & 56.93 & 88.54 & 70.74\rlap{$\text{}^{\mathsection}$}& 74.35\rlap{$\text{}^{\mathsection}$} & 91.27\\
     & Image & 65.73 & 66.91 & 94.66 & 76.26 & 80.11 & 96.08 \\
     & Multi& 64.21 & 61.63  & 91.43 & 54.61\rlap{$\text{}^{\mathsection}$} & 62.16\rlap{$\text{}^{\mathsection}$} & 92.52\\\hline
    \multirow{3}{*}{\shortstack{w/o Prompt}}& Text  & 63.86 & 58.00\rlap{$\text{}^{\ddagger}$} & 89.60\rlap{$\text{}^{\ddagger}$} & 69.99 & 73.82 & 91.65 \\
     & Image & 66.53\rlap{$\text{}^{\dagger \ddagger}$} & 67.56\rlap{$\text{}^{\dagger \ddagger}$} & 94.42 & 76.07 & 80.14 & 96.58\rlap{$\text{}^{\dagger \ddagger}$} \\
     & Multi& 64.92\rlap{$\text{}^{\ddagger}$} & 62.50\rlap{$\text{}^{\ddagger}$} & 91.81\rlap{$\text{}^{\ddagger}$} & 54.20 & 61.79 & 92.93\rlap{$\text{}^{\ddagger}$}\\\hline
    \end{tabular}}
\caption{\label{tab:ablation}Ablation Studies. ${\dagger}$, ${\ddagger}$, and ${\mathsection}$ indicate statistically significant improvements over MARVEL-ANCE w/o $\text{CLIP Pretraining}^{\dagger}$, $\text{MARVEL-ANCE w/o MS MARCO Pretraining}^{\ddagger}$ and $\text{MARVEL-ANCE w/o Prompt}^{\mathsection}$.}
\end{table*}
\section{Evaluation Result}
In this section, we first evaluate the performance of MARVEL and conduct ablation studies. Then, we explore the effectiveness of different visual and language model fusion methods and analyze the role of visual module adaption pretraining in MARVEL. Some case studies are shown in Appendix~\ref{app:case}.

\subsection{Overall Performance}
The multi-modal retrieval performance of MARVEL and baseline models is shown in Table~\ref{tab:overall}.
Besides retrieval performance, we also compared retrieval efficiency in Appendix \ref{app:time}. 

Overall, MARVEL significantly outperforms baseline models on all datasets by achieving more than 2\% improvements on both datasets, demonstrating its advantages in multi-modal retrieval tasks. Compared with text retrieval models, MARVEL improves their performance, showing that the image features are crucial in the multi-modal retrieval task. Furthermore, these universal multi-modal dense retrievers, UniVL-DR and MARVEL, outperform divide-and-conquer models by alleviating the modality fusion problem~\cite{liu2023univldr}. Compared with our main baseline UniVL-DR, MARVEL encodes queries and multi-modal documents using a universal encoder. Experimental results show that MARVEL significantly improves the retrieval effectiveness of UniVL-DR on both datasets, demonstrating its effectiveness in bridging the modality gap between images and texts.

\subsection{Ablation Study}\label{sec:ablation}
As shown in Table~\ref{tab:ablation}, we conduct ablation studies to explore the role of different modules of MARVEL in multi-modal retrieval. More ablation studies are shown in Appendix~\ref{app:ablation}.

In the comparison between MARVEL and MARVEL w/o CLIP Pretraining, pretraining the visual understanding module shows its effectiveness by improving the performance on single/multi-modal retrieval tasks. It shows that the image-caption alignment relations provide some opportunities to adapt the visual module to the language model via pretraining. Subsequently, MARVEL also outperforms MARVEL w/o MS MARCO Pretraining, especially on the text retrieval task. It demonstrates that MARVEL can broaden the advantage of text relevance modeling to the multi-modal retrieval task. To unify the multi-modal encoding, MARVEL follows previous work~\cite{hannan2020manymodalqa} uses prompt tokens to indicate the start and end positions of encoded image features (Eq.~\ref{eq:imagemap}), aiming to distinguish the image features from text token embeddings. These image prompt tokens bring light improvements, illustrating their roles in multi-modal document representation.

\subsection{Retrieval Effectiveness of Different Visual-Language Fusion Methods}\label{sec:fusion}
In this experiment, we show the retrieval effectiveness of MARVEL on the WebQA dataset by using different modality fusion and finetuning methods.

\begin{table}[ht]
 \centering

 \resizebox{\linewidth}{!}{
    \begin{tabular}{l|c|cc}
    \hline
    \textbf{Method}    & \textbf{Modality} & \textbf{MRR@10} & \textbf{Rec@100} \\
    \hline
        & Text & 51.75 & 84.37 \\
        CLIP-Sum & Image & 60.61 & 94.84 \\
        & Multi & 48.83 & 86.43 \\\hline
                        & Text    & 51.84  & 85.06   \\

    T5-CLIP (Sum)    & Image  & 58.09  & 93.13   \\
                        & Multi  & 35.03  & 79.00   \\
    \hline
                        & Text    & 48.71  & 81.78  \\
    T5-CLIP (Concat) & Image    & 37.20  & 81.14   \\
                        & Multi  & 25.19  & 62.77   \\
    \hline
                            & Text    & 54.28  & 85.80    \\
    T5-CLIP (Plugin) & Image    & 60.81  & 93.55   \\
                        & Multi  & 55.58  & 88.50   \\
    \hline
    \end{tabular}}
 \caption{\label{tab:fusion}Retrieval Performance of the Models using Different Visual-Language Fusion Methods. T5-CLIP (Sum/Concat) is similar to previous work~\cite{liu2023univldr}, which only replace the image caption encoder with T5-ANCE. The CLIP-Sum model is the CLIP-DPR model from previous work~\cite{liu2023univldr}. All models are trained with in-batch negatives. MRR@10 is used to evaluate the retrieval effectiveness of all models.}
\end{table}

\textbf{Modality Fusion.} Three kinds of visual-language fusion methods are compared in our experiments, including Sum, Concat and Plugin. For Sum and Concat methods, we encode the captions and image features separately as embeddings, then sum or concatenate these embeddings, followed by joint training of T5-ANCE and CLIP models with in-batch negatives. We show the experimental results in Table~\ref{tab:fusion}. MARVEL's visual module plugin method outperforms other fusion methods. This highlights the effectiveness of utilizing pretrained attention heads of language models for extracting image semantics and fostering deeper interactions between image and text inputs. Our plugin method proves instrumental in mitigating the modality gap between texts and images, enabling MARVEL to better represent image documents by jointly modeling image captions and features.

Different from~\citet{liu2023univldr}, we use T5-ANCE and CLIP as the text and image encoders, respectively. These models have different architectures and are pretrained on text retrieval and image-caption matching tasks. 
The multi-modal retrieval performance of CLIP-Sum decreases when we encode the image caption with a stronger retrieval model T5-ANCE (T5-CLIP-Sum) instead of CLIP. It demonstrates that incorporating an additional visual module into a well-trained dense retrieval model is still challenging for multi-modal retrieval. Notably, MARVEL provides a promising way to enable the image understanding ability of dense retrieval models by using the visual module plugin modeling method.



\begin{figure}[t]
    \centering
    \subfigure[Attention Weights.] { \label{fig:embed:weight} 
    \includegraphics[width=0.48\linewidth]
    {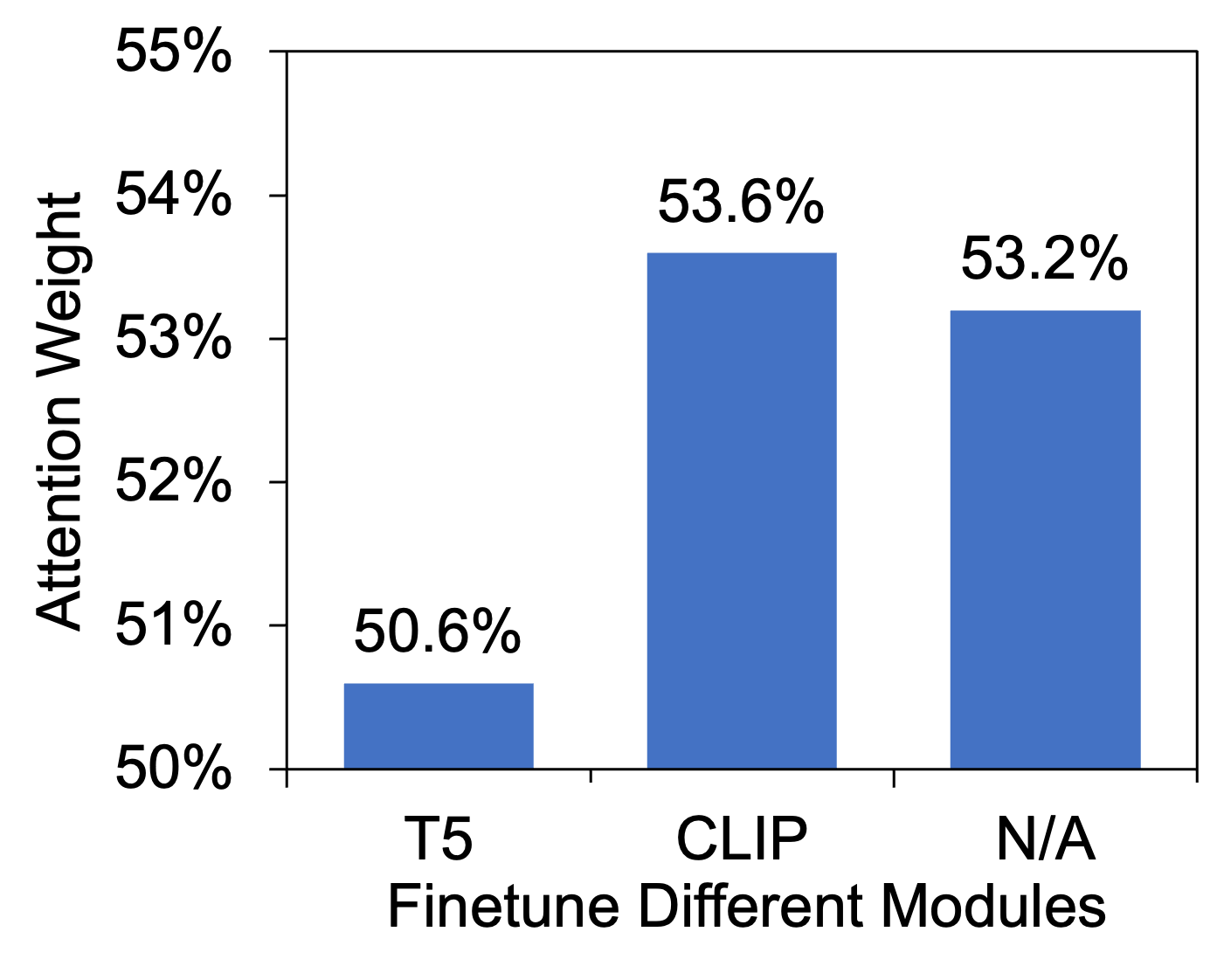}}  
    \subfigure[Attention Entropy.] { \label{fig:embed:entropy} 
    \includegraphics[width=0.48\linewidth]{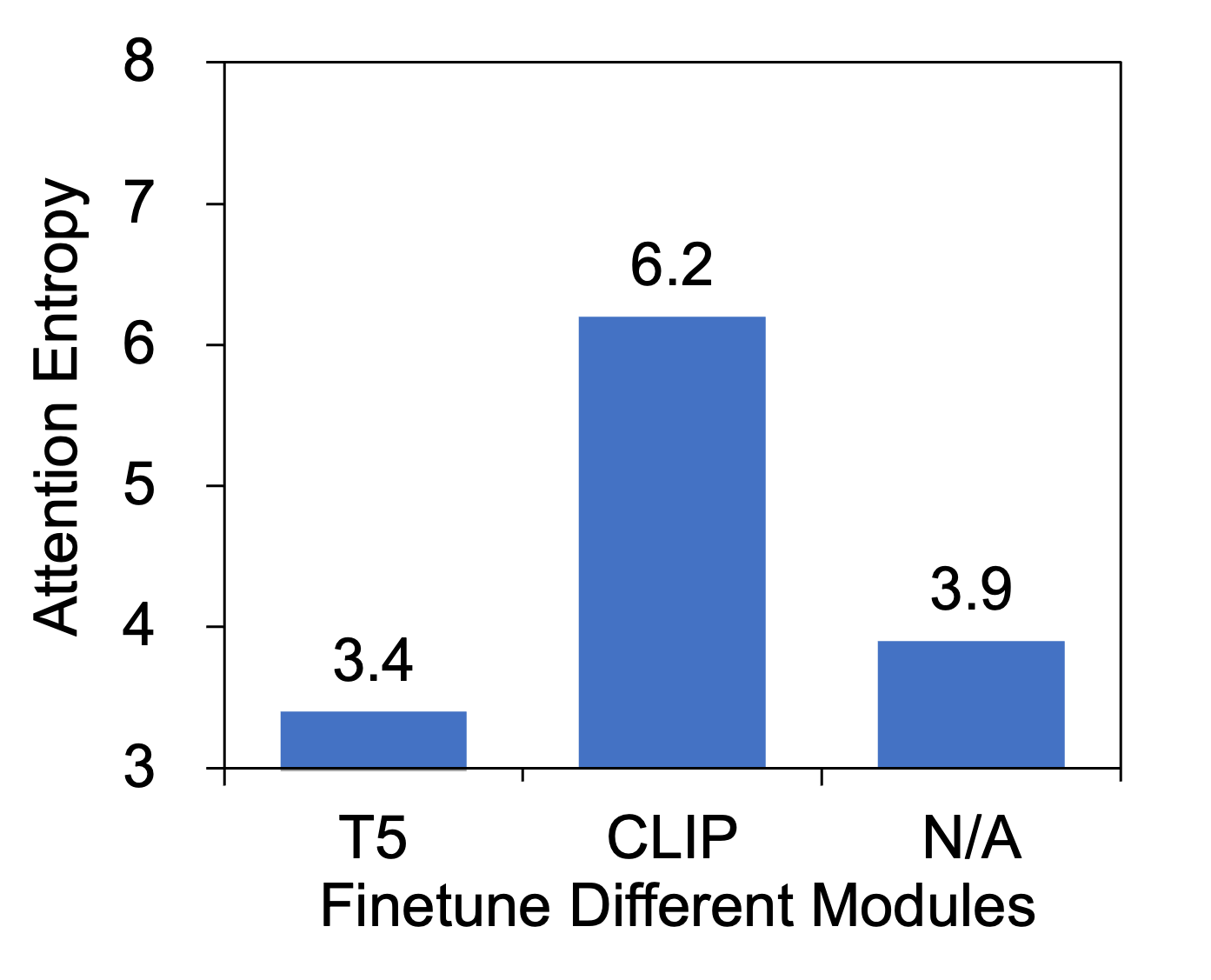}}

    \caption{Attention Distribution of MARVEL-ANCE. The attention weights of image features are shown in Figure~\ref{fig:embed:weight}. And the attention weight entropy of image captions and features is shown in Figure~\ref{fig:embed:entropy}.}
    \label{fig:embed}
\end{figure}
\textbf{Finetuning Strategies.} We then show the effectiveness of different finetuning strategies. In this experiment, we individually finetune the language model (T5) and visual module (CLIP) to show the changes of attention distributions and analyze the behaviors of different finetuning strategies.

As shown in Figure~\ref{fig:embed}. The attention scores are calculated by cross attentions from the decoder to the encoder module of T5. We first show the attention weight distribution of image features in Figure~\ref{fig:embed:weight}. 
When we only finetune the language model, the attention heads tend to allocate more balanced attention weights between image features and captions, helping to adapt the visual module in the language model. 
On the other hand, the image features win more attention weights when the CLIP module is finetuned. However, as shown in Figure~\ref{fig:embed:entropy}, only finetuning the CLIP module shows a scattered attention weight mechanism than other models, which misleads the T5-ANCE to capture more important information from encoded representations of documents. All these phenomena demonstrate the necessity of the training strategies of MARVEL, which pretrain visual module for adaption and only finetune the language model for multi-modal retrieval. 
In addition, we show the retrieval effectiveness with different finetuning methods in Appendix~\ref{app:finetuning}.

\begin{table*}[ht]
    \centering
    \resizebox{\linewidth}{!}{
    \begin{tabular}{l l}
    \hline

    \textbf{Figure} & \multicolumn{1}{c}{\textbf{Text}}\\\hline
    
    \multirow{8}[0]{*}{\begin{minipage}[b]{0.45\columnwidth}
		\raisebox{5mm}{\includegraphics[height=27mm,width=36mm]{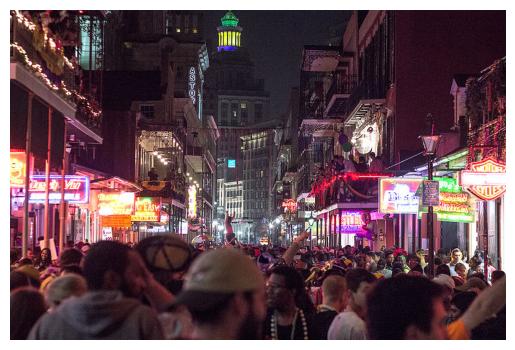}} 
	\end{minipage}} &  \small{\textbf{Manual Caption:} Mardi Gras Bourbon Street 2015 Bourbon Street, New Orleans, during Mardi Gras }\\[0.15cm] 

    &  \small{\textcolor[rgb]{0,0,1}{Nearest Tokens:} [``kehr'',``voted'', \textcolor[rgb]{0.7,0.3,0.3}{``brightness''}, ``event'', ``city'', ``venue'', ``local'', ``pub'', ``bounce'', ``island'', ``ferry'', }\\
    & \small{``keto'', ``Ice'', \textcolor[rgb]{0.7,0.3,0.3}{``residents''}, ``lighting'', \textcolor[rgb]{0.7,0.3,0.3}{``store''}, ``lights'', ``banks'', ``Lake'', ``impacted'', ``lively'', ``drinks'', ``eye'']}\\[0.1cm]

    &  \small{\textcolor[rgb]{0,0,1}{Nearest Tokens w/o CLIP Pretraining:} [``OUG'', ``7,000'', ``ban'', ``CU'', ``edited'', ``ition'', ``Pop'', ``imprisonment'',}\\
    & \small{``O'', ``militari'', ``immuno'', ``Ton'', ``reset'', ``États'', ``concise'', ``Arbeits'', ``IN'', ``Hi'', ``RAM'',}\\
    & \small{``Hello'', ``stocked'', ``charged'', ``institu''] }\\[0.05cm]
    \hline
    
    \multirow{8}[0]{*}{\begin{minipage}[b]{0.45\columnwidth}
		\raisebox{5mm}{\includegraphics[height=27mm,width=36mm]{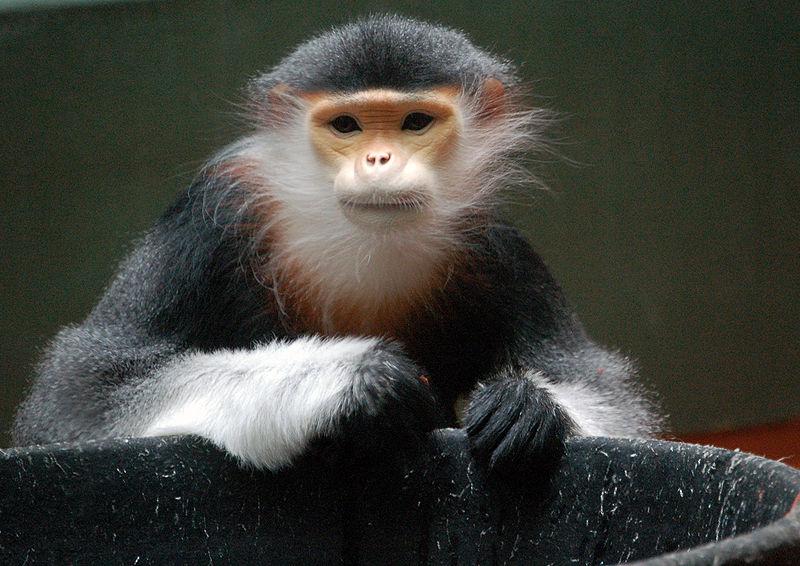}} 
	\end{minipage}} & \small{\textbf{Manual Caption:} Red-shanked Douc at the Philadelphia Zoo}\\[0.15cm]

    & \small{\textcolor[rgb]{0,0,1}{Nearest Tokens:} [``Whale'', ``endangered'', ``horn'', ``bird'', ``goat'', \textcolor[rgb]{0.7,0.3,0.3}{``animals''}, \textcolor[rgb]{0.7,0.3,0.3}{``wildlife''}, ``mammals'', ``whale'', ``Gib'',}\\
    & \small{``Elephant'', ``Savannah'', ``dach'', ``birds'', \textcolor[rgb]{0.7,0.3,0.3}{``creatures''}, \textcolor[rgb]{0.7,0.3,0.3}{``Wildlife''}, ``lois'', ``biomass'', ``limb'', \textcolor[rgb]{0.7,0.3,0.3}{``Creatures''}]} \\[0.1cm]

    &  \small{\textcolor[rgb]{0,0,1}{Nearest Tokens w/o CLIP Pretraining:} [``RAM'', ``bilingual'', ``MOD'', ``native'', ``recognizable'', ``Graphic'', ``charged'',}\\
    & \small{``ordentlich'', ``gray'', ``suffisamment'', ``colorful'', ``clar'', ``haunt'', ``riad'', ``CM'', ``ammunition'', ``ordre'', ``thetic'',}\\
    & \small{``Hi'', ``auftrag'', ``he'', ``ban'', ``sets'', ``7,000'', ``representation'']}\\[0.05cm]
   
     \hline
    
    \multirow{8}[0]{*}{\begin{minipage}[b]{0.45\columnwidth}
		\raisebox{5mm}{\includegraphics[height=27mm,width=36mm]{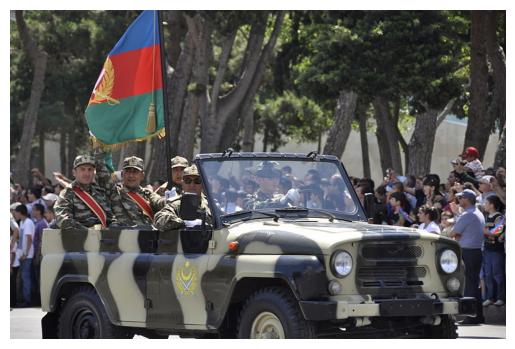}} 
	\end{minipage}} & \small{\textbf{Manual Caption:} Military parade in Baku on an Army Day28 Military parade in Baku on an Army Day} \\[0.15cm] 
    & \small{\textcolor[rgb]{0,0,1}{Nearest tokens:} [\textcolor[rgb]{0.7,0.3,0.3}{``vehicles''}, ``Fahrzeug'', ``territories'', \textcolor[rgb]{0.7,0.3,0.3}{``flag''}, ``chemical'', ``replies'', ``migrants'', ``parliament'', ``bikes'',}\\
    & \small{\textcolor[rgb]{0.7,0.3,0.3}{``militari''}, ``equipment'', ``République'', ``troops'', ``clothing'', ``gear'', ``prisoners'', ``machinery'', ``tribe'', ``vorgesehen'']}\\[0.1cm]

    &  \small{\textcolor[rgb]{0,0,1}{Nearest Tokens w/o CLIP Pretraining:} [ ``împreună'', ``RAM'', ``ened'', ``troupe'', ``Compet'', ``sie'', ``own'', ``RGB'',}\\
    & \small{ ``Ha'',``operation'', ``arbeit'', ``enforcement'', ``Cor'', ``EU'', ``LCD'', \textcolor[rgb]{0.7,0.3,0.3}{``countries''}, ``SO'', ``institu'', ``grief'', ``limbi'', }\\
    & \small{``default'', ``16'', ``raum'', ``haunt'', ``unanimous''] }\\[0.05cm]
    \hline

    \multirow{8}[0]{*}{\begin{minipage}[b]{0.45\columnwidth}
		\raisebox{5mm}{\includegraphics[height=27mm,width=36mm]{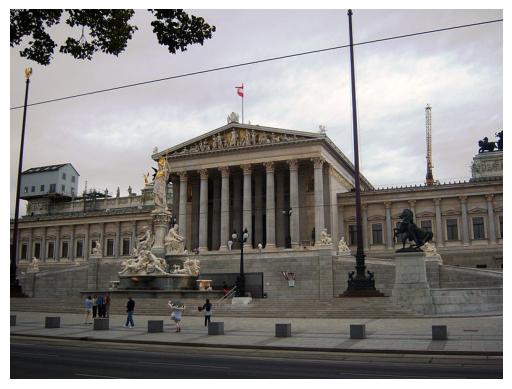}} 
	\end{minipage}} & \small{\textbf{Manual Caption:}Parlament Wien Austria, Vienna, Austrian Parliament Building }\\[0.15cm]
     & \small{\textcolor[rgb]{0,0,1}{Nearest Tokens:} [``Schloss'', ``funds'', ``Statut'', ``furniture'', ``Albany'', ``structure'', ``palace'', ``Capitol'', ``statute'', ``locul'',}\\
     & \small{``headquarters'', ``occupie'', ``structures'', \textcolor[rgb]{0.7,0.3,0.3}{``legislature''}, ``cinéma'', \textcolor[rgb]{0.7,0.3,0.3}{``legislation''}, \textcolor[rgb]{0.7,0.3,0.3}{``governmental''}, ``Argentin'',``vederea'']}\\ [0.1cm]

      &  \small{\textcolor[rgb]{0,0,1}{Nearest Tokens w/o CLIP Pretraining:} [ ``militari'', ``reset'', ``Ton'', ``shrine'', ``commands'', ``hi'', ``împreună'', ``Achtung'',}\\
      & \small{``genug'', ``shake'', ``RAM'', ``iconic'', ``committed'', \textcolor[rgb]{0.7,0.3,0.3}{``département''}, ``colo'', ``Hi'', ``Sammlung'', ``pop'', ``1951'', ``ban'',}\\
      & \small{``russia'', ``Color'', ``vivid'', ``HM'', ``arbeit'', ``default''] }\\[0.05cm]
      \hline



\end{tabular}
}
    \caption{The Nearest Tokens of Image Features. We randomly select five image documents, encode these image features using the visual module of MARVEL and MARVEL w/o CLIP Pretraining, and then show the nearest tokens of the encoded image features. The tokens related to the semantics of the image document are \textcolor[rgb]{0.7,0.3,0.3}{highlighted}.}
    \label{tab:image_nearest_token_cases}%
\end{table*}%
\subsection{Effectiveness of Visual Module Adaption Pretraining}\label{sec:image_prompt}
We show some cases in Table~\ref{tab:image_nearest_token_cases} to show the effectiveness of the visual understanding module by verbalizing encoded image features. 
Besides, more experiments about the encoded image features are shown in Appendix~\ref{app:replacement}. 

We randomly select four image documents of different topics and represent the encoded image feature with some tokens to verbalize the image semantics. Specifically, we first use the visual plugin modules of MARVEL and MARVEL w/o CLIP Pretraining to encode the image features. Then, to show the semantics of the encoded image features, we utilize cosine similarity to find the $k$-NN tokens for each encoded image feature. Finally, the tokens with the highest score are used to represent the semantics of the encoded image features.

For the first two examples, MARVEL w/o CLIP Pretraing learns more similar representations for both image documents. The related word tokens of these image documents contain lots of same tokens, such as ``7,000'', ``Hi'', and ``RAM'', which are unrelated to the semantics of the image documents. On the contrary, our MARVEL model can learn more similar semantics to both image documents. Specifically, MARVEL verbalizes the first image document using the words ``brightness'', ``resident'' and ``store'', which are related to the image description of ``Bourbon Street''. Additionally, MARVEL captures the semantics of ``animals'', ``wildlife'' and ``creatures'' in the second image document. 
The next two instances show the effectiveness of MARVEL in learning more fine-grained semantics of the image documents by verbalizing the image documents with more related words, such as ``militari'', ``vehicle'', ``flag'', ``legislatur'', and ``government''. 
Overall, compared with the MARVEL w/o CLIP Pretraining model, MARVEL has the ability to learn more effective representations that are closer to the semantics of the images, demonstrating the important role of MARVEL's visual module pretraining strategy in adapting the visual understanding module for dense retrievals.




\section{Conclusion}
This paper proposes \textbf{M}ulti-mod\textbf{A}l \textbf{R}etrieval via \textbf{V}isual modul\textbf{E} p\textbf{L}ugin (MARVEL). MARVEL integrates a visual plugin module with a well-trained dense retriever and pretrains the visual module with image-caption contrastive training for adaption. Our MARVEL model achieves state-of-the-art on all benchmarks by unifying the multi-modal document encoding and alleviating the modality gap between images and texts. 
\section*{Limitations}
Even though MARVEL shows strong effectiveness in the multi-modal retrieval task, there are some limitations in our work. Existing multi-modal retrieval systems still highly depend on the semantics of image caption instead of the image understanding ability of the visual module. In this case, MARVEL pretrains the visual understanding module but achieves limited improvements. Building an effective visual understanding module is crucial for the multi-modal retrieval task.

\section*{Acknowledgments}
This work is partly supported by the Natural Science Foundation of China under Grant (No. 62206042, No. U23B2019, No. 62137001, and No. 62272093), the Joint Funds of Natural Science Foundation of Liaoning Province (No. 2023-MSBA-081), and the Fundamental Research Funds for the Central Universities under Grant (No. N2416012).

\bibliography{citation}
\clearpage
\newpage
\appendix
\section{Appendix}

\subsection{License}
We show the licenses of the datasets that we use. WebQA uses CC0-1.0 license, while ClueWeb22 shows its terms of use at website\footnote{\url{https://lemurproject.org/clueweb22/}}. 
All these licenses and agreements permit the use of their data for academic purposes.

\subsection{Experimental Details of MARVEL Pretraining Data}\label{app:pretrain}
In this subsection, we introduce the experimental details to process the pretraining data.

To pretrain the visual module in MARVEL, we collect the image-caption pairs from the ClueWeb22 dataset. We retain the English pages, extract the content within the image tag and use the image and alt-text to construct the image-caption pair. 
To ensure the quality of the pretraining dataset, following LAION-400M~\cite{schuhmann2021laion}, we use CLIP to calculate the embeddings of images and captions and compute the cosine similarity between the two embeddings. Subsequently, we discard all samples with a cosine similarity lower than 0.3. 
The pretraining dataset contains 1.6M image-caption pairs, and we randomly select 10,000 pieces of data as the development set and use the rest for the pretraining visual module.



\begin{table}
 \centering

   \resizebox{0.48\textwidth}{!}{
    \begin{tabular}{c|c|ccc}
    \hline
    \textbf{Finetune}                & \textbf{Modality} & \textbf{MRR@10} & \textbf{NDCG@10} & \textbf{Rec@100} \\
    \hline
                  & Text  & 64.89  & 58.71   & 89.98   \\
    CLIP \& T5    & Image  & 64.36  & 65.41   & 94.19   \\
                  & Multi  & 64.37  & 61.77   & 91.78   \\
    \hline
                & Text    & 64.72  & 58.88   & 90.26   \\
    T5          & Image  & 66.12  & 67.49   & 95.12   \\
                & Multi  & 65.15  & 62.95   & 92.40   \\
    \hline
                & Text    & 48.38  & 41.63    & 75.11  \\
    CLIP        & Image    & 56.28  & 56.17   & 87.67   \\
                & Multi  & 49.22  & 45.80   & 80.42   \\
    \hline
                & Text    & 48.38  & 41.39   & 74.57    \\
    N/A         & Image & 55.09  & 54.99   & 87.26   \\
                & Multi  & 48.12  & 44.75  & 79.69   \\
    \hline
    \end{tabular}}
 \caption{\label{tab:fix}The Retrieval Performance with Different Training Strategies. We freeze each module of MARVEL-ANCE to explore the benefits of training between different modules.}
\end{table}
\subsection{Retrieval Effectiveness of Different Finetuning Strategies}\label{app:finetuning}
In this experiment, we show the performance of single/cross and multi-modal retrieval tasks with different finetuning strategies.

As shown in Table~\ref{tab:fix}, finetuning the CLIP module indeed improves the retrieval performance of the whole frozen model, especially in the image retrieval task. This observation shows that multi-modal training signals are effective to benefit the capability of visual modules. When we only tune the parameters of T5, MARVEL-ANCE achieves significant improvements over the frozen model, showing the language model's strong ability to adapt the visual module to the dense retriever. Nevertheless, the fully finetuned model decreases the retrieval performance of MARVEL-ANCE that only finetunes T5. It shows the necessity of the pretraining-and-then-finetuning strategy of MARVEL, which pretrains the visual understanding module for adaption and finetunes the language model for multi-modal retrieval.

\subsection{More Details of ClueWeb22-MM}\label{app:clueweb}
\begin{table}
\centering

\resizebox{\linewidth}{!}{
\begin{tabular}{l|c c c c c}
\hline  
\textbf{Data Type} & \textbf{Median} & \textbf{Average} & \textbf{Max}& \textbf{Min} \\
\hline
\textbf{Queries} & 8.0  & 9.9 & 245.0 & 1.0 \\
\hline
\textbf{Text Documents}& 52.0  & 127.8 & 1121183.0 & 1.0 \\
\hline
\textbf{Image Captions}& 6.0  & 8.1 & 998.0 & 1.0 \\
\hline

\end{tabular}}
\caption{\label{tab:cluewebnum}Length Statistics of Queries, Text Documents and Image Captions in ClueWeb22-MM Dataset.}
\end{table}
\begin{table}
\small 
    \centering
    
        \begin{tabular}{l| c}
            \hline  
            \textbf{Range of Image Sizes} & \textbf{Number} \\
            \hline  
            Height or Width $\geq$ 1024 & 23.8k \\
            \hline  
            Height and Width $\geq$ 1024 & 7.4k \\
            \hline  
            Height or Width $\geq$ 512 & 81.9k \\
            \hline  
            Height and Width $\geq$ 512 & 43.9k \\
            \hline  
            Height or Width $\geq$ 256 & 234.6k \\
            \hline  
            Height and Width $\geq$ 256 & 170.2k \\
            \hline  
        \end{tabular}    
    \caption{\label{tab:cluewebimagedis}Image Size Distribution of ClueWeb22-MM.}
    \normalsize 
\end{table}
\begin{table*}[!ht]

\small
  
\begin{tabular}{p{0.95\linewidth}}
\hline

\makecell[c]{\textbf{Queries with the Text Document as Label}}
\tabularnewline 
\hline
\textbf{Query:} Chinese Dragons — Facts, Culture, Origins, and Art
\\

\textbf{Text Document:} Live updates on China travel restrictions for 2022.  Learn more Home Chinese Culture Traditional Chinese Clothes  Chinese Dragons — Facts, Culture, Origins, and Art Written by  Mike Ho Updated Dec. 14, 2021 Chinese dragons are powerful and benevolent symbols in Chinese culture, with supposed control over watery phenomenon, e.g. summoning rain during a drought. Dragons are everywhere in China — in legends, festivals, astrology, art, names, and idioms.
\tabularnewline 
\hline

\textbf{Query:} How to manage partitions with the Disk Management tool, in Windows | Digital Citizen
\\

\textbf{Text Document:} Disk Management A new window should pop up, listing the drive letter of the partition. Click or tap Change and, in the next window, select the new drive letter you wish to assign to it. Then, click or tap OK.
\tabularnewline
\hline

\textbf{Query:} here’s a small-batch peanut butter oatmeal cookie recipe for you
\\

\textbf{Text Document:} You are here:  Home / Recipes /  Small-batch Peanut Butter Oatmeal Cookies Small-batch Peanut Butter Oatmeal Cookies 02/21/19  |  Cookies , Desserts ,  Recipes ,  Small-batch Dessert These Small-batch Peanut Butter Oatmeal Cookies are the perfect cookie hybrid. They’re rich and peanut buttery, bendy and chewy, and the best of both worlds. A few weeks ago, I posted these (AMAZING)  Peanut Butter Oatmeal Cookies . It was a big-batch recipe meant for sharing and freezing, so I promised that I’d add a small-batch version ASAP for those of you who are here for small-batch desserts. So here we go. Let’s make a cute little batch of Peanut Butter Oatmeal Cookies and share with no one.
\tabularnewline 
\hline

\textbf{Query:} What foods increase uric acid
\\

\textbf{Text Document:} Vegetables and legumes that increase uric acid Legumes such as lentils, chickpeas or beans are known for their purine content, so their intake should be limited to only once or twice a week if you have high uric acid. Other vegetables that should be eaten in moderation are asparagus, mushrooms, cauliflower, spinach, radishes and leeks... Other foods that increase uric acid Other foods  that increases uric acid  and should be avoided are: All kinds of alcoholic beverages , especially beer and wine. Carbonated beverages, sugar-laden soft drinks and packaged juices. Avoid cooking with brewer's yeast...
\tabularnewline
\hline

\makecell[c]{\textbf{Queries with the Image Document as Label}}
\tabularnewline 
\hline

\makecell{\begin{minipage}{0.3\textwidth} 
  \includegraphics[height=20mm,width=30mm]{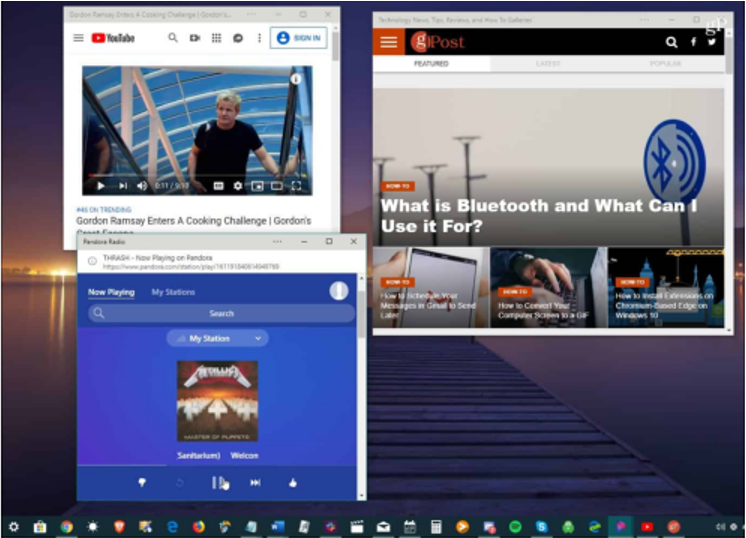}
\end{minipage}
\begin{minipage}{0.7\textwidth} 
  \textbf{Query:} Use Web apps With the New Chromium Edge on Windows 10 \\
  \textbf{Image Caption:} Web Apps Running Chromium Edge
\end{minipage}}%
\tabularnewline 
\hline

\makecell{
\begin{minipage}{0.3\textwidth} 
  \includegraphics[height=20mm,width=30mm]{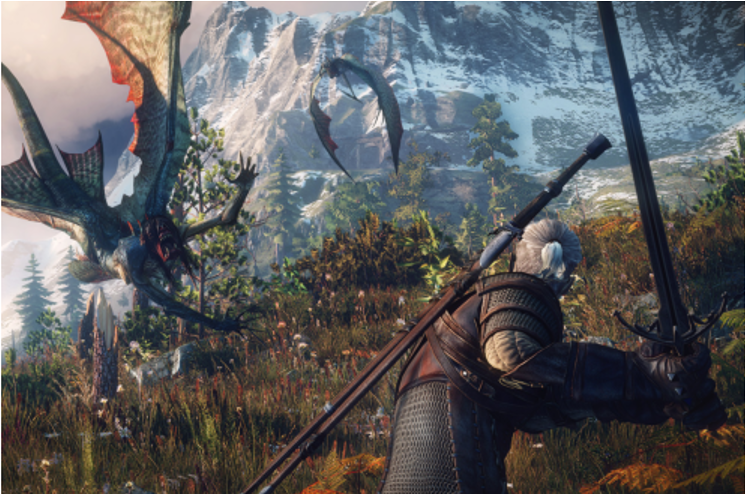}
\end{minipage}
\begin{minipage}{0.7\textwidth}\textbf{Query:} What are Runestones In Witcher 3?\\
\textbf{Image Caption:} Witcher 3 best runewords
\end{minipage}}%
\tabularnewline 
\hline

\makecell{
\begin{minipage}{0.3\textwidth} 
  \includegraphics[height=20mm,width=30mm]{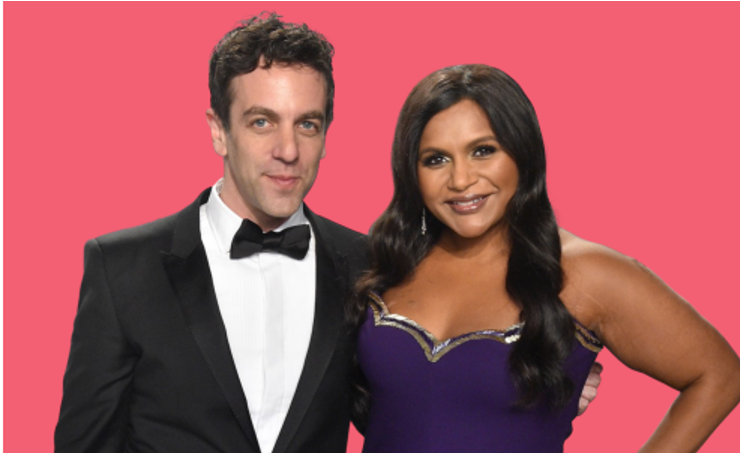}
\end{minipage}
\begin{minipage}{0.7\textwidth}\textbf{Query:} Everything We Know About Mindy Kaling and BJ Novak's\\ Relationship—Including Sweet Details from Her Book\\
\textbf{Image Caption:} mindy-kaling-bj-novak-removebg
\end{minipage}}%
\tabularnewline 
\hline

\makecell{
\begin{minipage}{0.3\textwidth} 
  \includegraphics[height=20mm,width=30mm]{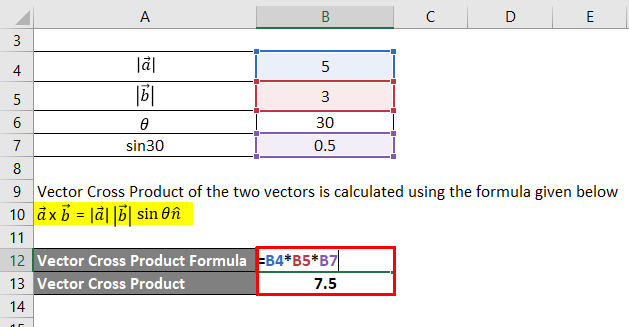}
\end{minipage}
\begin{minipage}{0.7\textwidth}\textbf{Query:} Vector Cross Product Formula Excel Template\\
\textbf{Image Caption:} Vector Cross Product Formula-1.2
\end{minipage}}%
\tabularnewline 
\hline

\end{tabular}
\caption{Examples of ClueWeb22-MM. We give practical examples of queries, image documents, and text documents.}
\label{tab:cluewebcase}
\end{table*}

\begin{table*}[ht]
  \centering
  \small
    \begin{tabular}{l|l|ccc|ccc}
    \hline
    \multirow{2}{*}{\textbf{Model}} & \multirow{2}{*}{\textbf{Modality}} & \multicolumn{3}{c|}{\textbf{WebQA}} & \multicolumn{3}{c}{\textbf{ClueWeb22-MM}}\\\cline{3-8}
    & &  \textbf{MRR@10} &\textbf{NDCG@10}  &  \textbf{Rec@100}  & \textbf{MRR@10}  &  \textbf{NDCG@10} & \textbf{Rec@100}\\ 
    \hline
    \multirow{3}{*}{\shortstack{MARVEL-ANCE}}& Text  & 64.72 & 58.88 & 90.26 & 71.73 & 75.40 & 92.29\\
     & Image & 66.12 & 67.49 & 95.12 & 77.57 & 81.34 & 96.50 \\
     & Multi& 65.15 & 62.95  & 92.40 & 55.19 & 62.83 & 93.16\\\hline
      \multirow{3}{*}{\shortstack{w/o Image Caption}}& Text  & 64.67 & 58.30 & 89.49 & 69.75 & 73.32 & 90.60\\
     & Image & 3.85 & 4.32 & 24.81 & 18.26 & 20.65 & 45.07 \\ 
     & Multi& 33.70 & 30.83  & 56.45 & 37.29 & 40.74 & 64.26 \\\hline
      \multirow{3}{*}{\shortstack{w/o Image Feature}}& Text  & 63.42 & 57.95 & 90.27 & 71.17 & 74.78 & 91.57\\
     & Image & 64.32 & 65.42 & 94.15 & 76.83 & 80.60 & 95.88 \\
     & Multi& 63.60 & 61.43  & 91.99 & 54.98 & 62.64 & 92.60\\\hline
    
    \end{tabular}
    \caption{\label{tab:appablation}Additional Ablation Study Results on MARVEL-ANCE.}
\end{table*}
\begin{table}
\centering
\small
  \resizebox{\linewidth}{!}{
\begin{tabular}{l|ccc}
\hline
\textbf{Model} & \textbf{MRR@10} & \textbf{NDCG@10} & \textbf{Rec@100} \\
\hline

MARVEL-DPR & 55.71 & 52.94 & 88.23 \\
w/ 1-NN Token & 38.80 & 35.89 & 73.59\\
w/ 5-NN Tokens & 42.39 & 39.27 & 75.04 \\
w/ Random Token & 37.73 & 35.34 & 71.92 \\
\hline
MARVEL-ANCE & 65.15 & 62.95  & 92.40 \\
w/ 1-NN Token  & 51.37 & 48.27 & 80.47 \\
w/ 5-NN Tokens  & 52.22 & 49.35 & 81.88 \\
w/ Random Token & 44.22 & 41.55 & 71.23 \\
\hline

\end{tabular}}
\caption{\label{tab:replacement}Multi-Modal Retrieval Performance of Different Image Feature Replacement Strategies. We conduct experiments on MARVAL-DPR and MARVAL-ANCE models by replacing the image features with the average of $k$-NN ($k$ Nearest Neighbour) word embeddings. The $k$ is set to 1 and 5.}
\end{table}


\begin{table*}[ht]
  \centering
  \small
    \begin{tabular}{l|l|ccc|c}
    \hline
    \multirow{2}{*}{\textbf{Setting}} & \multirow{2}{*}{\textbf{Model}} & \multicolumn{3}{c|}{\textbf{Encoding Time (ms)}} &\textbf{Retrieval Time (ms)}\\\cline{3-6}
    & &  \textbf{Query} & \textbf{Img Doc} & \textbf{Text Doc} & \textbf{Query}\\ 
    \hline
    \multirow{8}{*}{\shortstack{Single Modality\\(Text Only)}} 
    & BM25 & - & 0.1 & 0.1 & 41.9 \\ 
    & DPR (Zero-Shot) & 1.8 & 2.7 & 4.7 & 48.4 \\ 
    & CLIP-Text (Zero-Shot) & 0.4 & 0.2 & 0.3 & 32.7 \\ 
    & Anchor-DR (Zero-Shot) & 2.2 & 2.0 & 2.0 & 47.7 \\ 
    & T5-ANCE (Zero-Shot) & 2.1 & 2.0 & 2.0  & 48.3 \\ 
    & BERT-DPR & 1.1 & 2.7 & 3.1 & 47.6 \\ 
    & NQ-DPR & 1.3 & 2.8 & 4.9 & 47.8 \\ 
    & NQ-ANCE & 1.3 & 2.7 & 4.7 & 48.1 \\ 
    \hline
    \multirow{3}{*}{Divide-Conquer} 
    & VinVL-DPR & 1.4 & 5.2 & 4.7 & 48.4 \\ 
    & CLIP-DPR & 0.4 & 0.8 & 0.3 & 32.8 \\ 
    & BM25 \& CLIP-DPR & 0.4 & 0.8 & 0.1 & 37.3 \\ 
    \hline
    \multirow{4}{*}{UnivSearch} 
    & CLIP & 0.4 & 0.8 & 0.3 & 32.8 \\ 
    & VinVL-DPR & 1.3 & 5.3 & 4.7 & 47.6 \\ 
    & UniVL-DR & 0.4 & 0.9 & 0.3 & 32.9 \\ 
    & MARVEL & 2.1 & 3.7 & 2.0 & 48.0 \\ 
    \hline
    \end{tabular}
    \caption{\label{tab:time} Retrieval Efficiency. We compare the encoding and retrieval times of different architectural models on the same device. 
    These models encode the multi-model documents and queries offline and construct the FlatL2 index using FAISS~\cite{johnson2019billion} for online retrieval.}
\end{table*}

To show the details of our ClueWeb22-MM dataset, we show the data collection, data processing, and data statistics in this subsection.

\textbf{Data Collection.} Following previous work in text retrieval~\cite{zhang2020selective,xie2023unsupervised}, we regard the anchor text as a query and assume that its linked web page is related to the query. Then we extract image documents and text documents from these anchor-linked web pages. To obtain image documents, we parse HTML to extract the content within the image tag, then use alt-text as image caption, and crawl the image features from the image URL.

\textbf{Data Processing.} Ensuring the quality and meaningfulness of the ClueWeb22-MM dataset, we conduct additional processing on the data to filter out noise data according to the quality of images and alt-texts. Concerning images, we retain data with image file extensions such as jpg/png/jpeg and discard samples with image URLs containing keywords, \textit{e.g.} ``logo'', ``button'', ``icon'', ``plugin'', or ``widget''. Besides, we exclude the example, which has empty alt-text, has ``no alt attribute'' and contains an alt-text that is shorter than 5.

To further guarantee the quality of the dataset, we use T5-ANCE~\cite{yu2023openmatch} to estimate the relevance between the anchor and its corresponding image document. We encode all captions of image documents using T5-ANCE, use the anchor texts as queries to retrieve the images and reserve the anchors that are ranked in the top 10. Finally, we respectively sample 10,000 queries to build the development and test set. The rest data are used for finetuning models, which contain 72,028 queries.

\textbf{Data Statistics.} We provide length statistics on queries, text documents, and image captions in Table~\ref{tab:cluewebnum} and present the image size distribution in Table~\ref{tab:cluewebimagedis}. Subsequently, as shown in Table~\ref{tab:cluewebcase}, we show eight examples to illuminate the ClueWeb22-MM dataset. These examples show that the anchor-document pairs are of high quality. Thus we can use them to build an evaluation benchmark for multi-modal retrieval.

\subsection{Additional Ablation Studies on MARVEL}\label{app:ablation}
We conduct additional ablation studies to explore the role of image captions and image features in the multi-modal retrieval task. 

As shown in Table~\ref{tab:appablation}, the relevance modeling between queries and image documents heavily depends on the image caption, which is also observed in previous work~\cite{liu2023univldr}. The image features contribute to approximately 1\% improvements in the image retrieval task, demonstrating the effectiveness of image features in helping the model better understand the image documents.

\subsection{Learned Semantics of Image Features}\label{app:replacement}
In this experiment, we explore the semantic information of image features encoded by the visual module on the WebQA dataset. 
During training MARVEL model, we map the encoded image features into the input space of T5-ANCE's word embeddings. 
We conduct several experiments by replacing the encoded image features with the embeddings of the nearest or random tokens.

As shown in Table~\ref{tab:replacement}, replacing encoded image features with $k$-NN token embeddings generally outperforms the retrieval model using randomly selected token embeddings. 
It demonstrates that the visual plugin module effectively maps image semantics in the input space of the language model, and the ability to keep growing with more token embeddings (5-NN). 
However, the retrieval performance significantly drops when employing $k$-NN token embeddings to replace the image features, compared to the MARVEL model. It demonstrates the role of encoded image features beyond the semantic representations of word embeddings. The encoded image features may act as a kind of prompt, encouraging language models to capture image semantics~\cite{merullo2022linearly}. 

\subsection{Retrieval Efficiency of MARVEL}\label{app:time}

In this section, we compare the retrieval efficiency of MARVEL with other baselines on the same device, as shown in Table \ref{tab:time}.

MARVEL follows the general dense retrieval framework~\cite{karpukhin2020dense} for efficient document retrieval. It encodes the entire corpus offline and constructs the document index using FAISS~\cite{johnson2019billion} for online searches. While the offline encoding time for queries and image/text documents in MARVEL is longer than that in UniVL and other baseline models, this encoding process does not impact retrieval efficiency.

When comparing the retrieval latency of these models in retrieving the top 100 relevant documents for each query, MARVEL’s retrieval time is comparable to its base model, T5-ANCE, which has a retrieval time of 48.3 ms. Furthermore, the retrieval time for document encoding using the CLIP model is less than that of models such as T5 and BERT. This demonstrates that the retrieval time is influenced by the dimensionality of the embeddings, and MARVEL's architecture does not introduce any additional retrieval latency.

\begin{figure*}[t]
\centering
    \subfigure[Top5 Multi-modal Documents Retrieved from WebQA.] { \label{fig:embed:zs-image} 
    \includegraphics[width=1\linewidth]{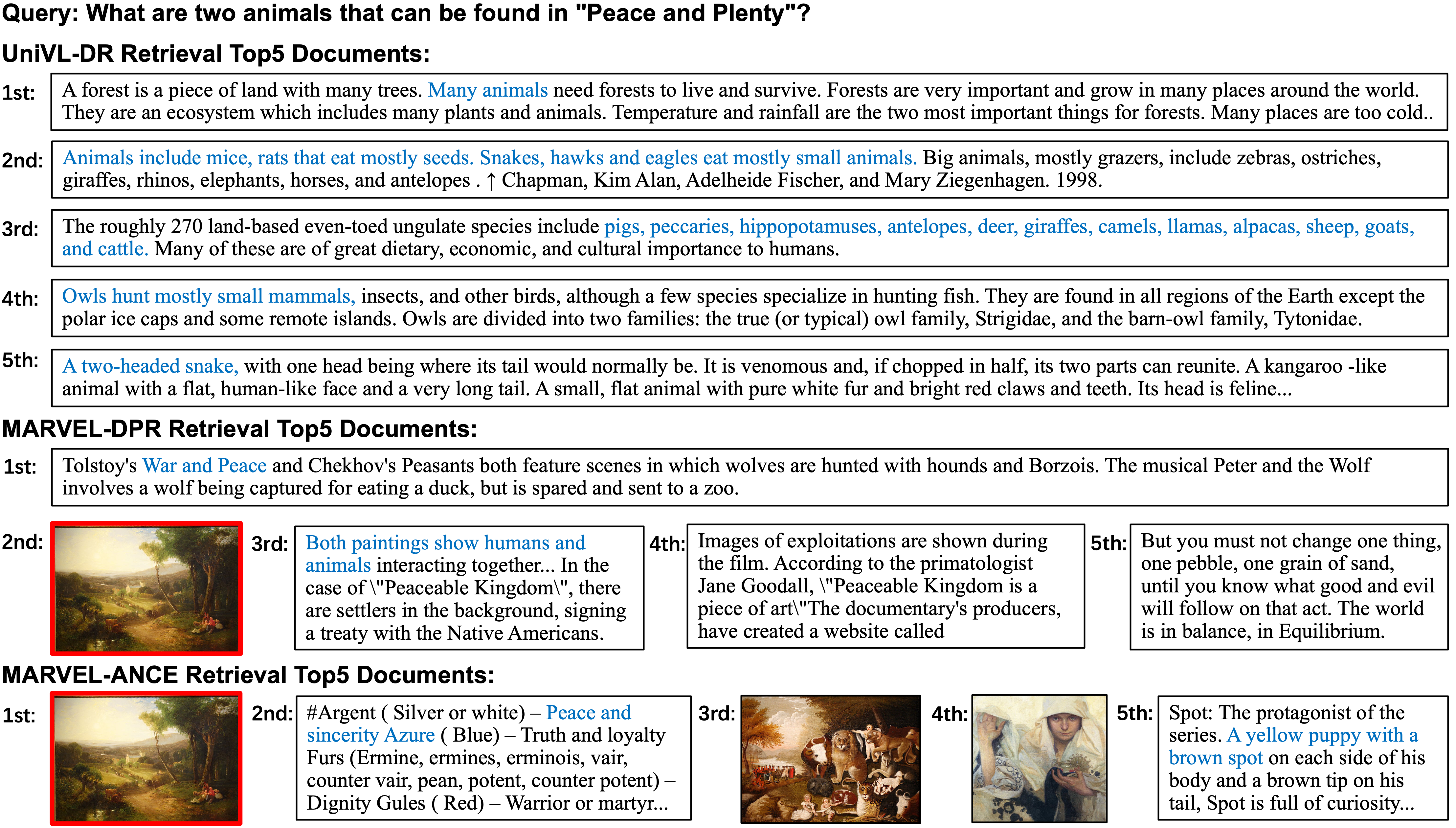}}
    
    \subfigure[Top5 Multi-modal Documents Retrieved from Clueweb22-MM.] { \label{fig:embed:zs-text} 
    \includegraphics[width=1\linewidth]{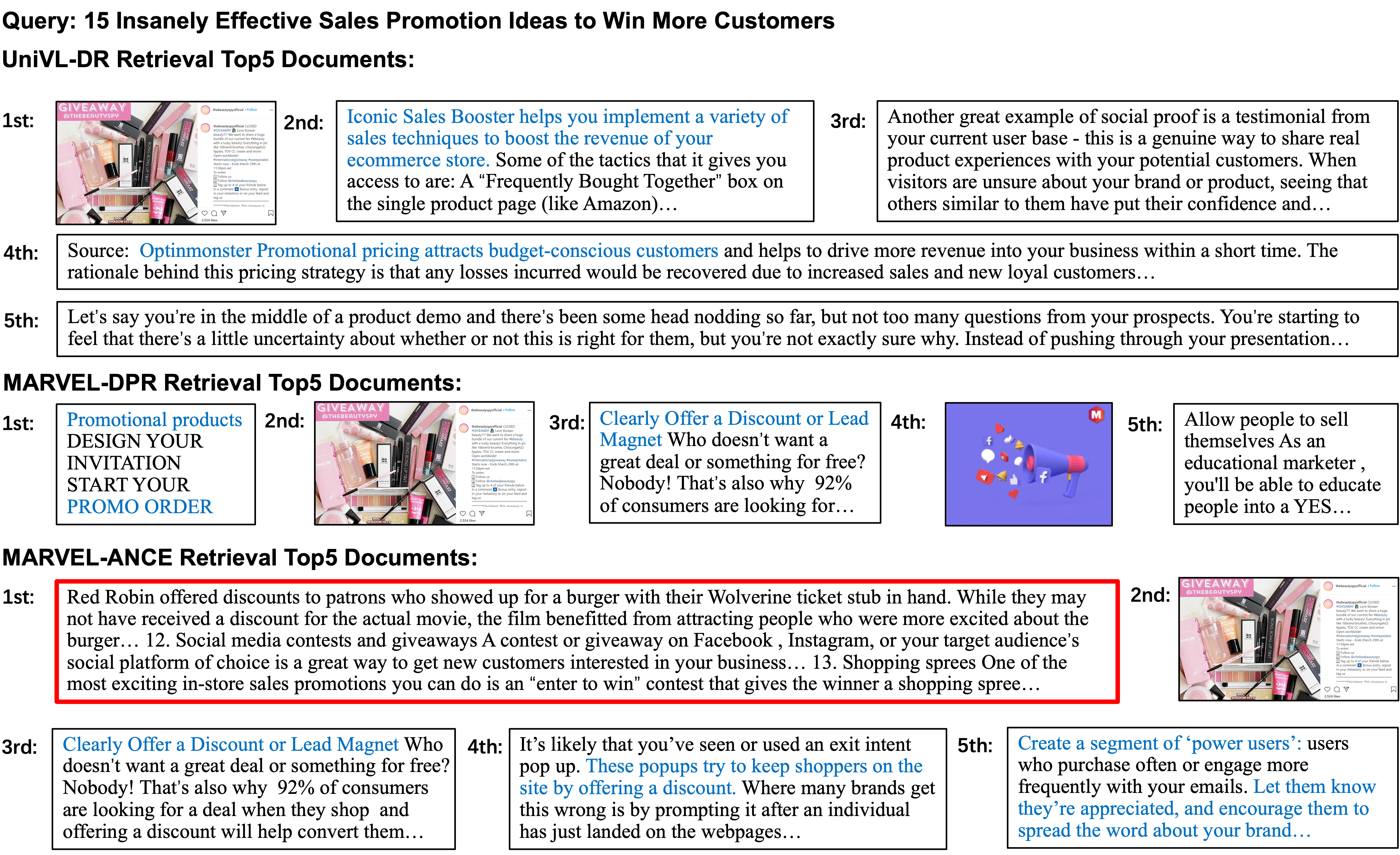}}
\caption{Case Studies. We present two cases from WebQA and ClueWeb22-MM and show the top5 retrieved multi-modal documents. The ground-truth documents and related content are highlighted in red and blue respectively.}
\label{fig:top5case}
\end{figure*}
\subsection{Case Studies}\label{app:case}
In Figure~\ref{fig:top5case}, we show two cases from WebQA and ClueWeb22-MM to study the multi-modal retrieval effectiveness of MARVEL. The top 5 documents retrieved by UniVL-DR, MARVEL-DPR, and MARVEL-ANCE are presented.

For the first case, UniVL-DR conducts shallow keyword matching and returns text documents that are related to ``animal'' and ``Peace'' mentioned in the query, which are unrelated to the query. In contrast, MARVEL can better understand that ``Peace and Plenty'' is a famous painting and retrieve more related images and text documents for users. 
In the second case, UniVL-DR, MARVEL-DPR, and MARVEL-ANCE all return documents related to ``promotion ideas''. Notable, MARVEL can better understand the user's query and return the expected modality. MARVEL-ANCE introduces a variety of sales promotion strategies rather than matching on ``promotion'' keywords. It shows the effectiveness of MARVEL in better fusing the retrieval results from different modalities, which thrives on universal multi-modal document encoding.

\end{document}